\documentstyle[aps]{revtex}
\begin{document}
\title{Traversable wormholes: the Roman ring}
\author{Matt Visser\cite{e-mail}}
\address{Physics Department, Washington University, St. Louis, 
         Missouri 63130-4899}
\date{gr-qc/9702043; 11 November 1996}
\maketitle
\begin{abstract}
In this brief report I introduce a yet another class of geometries
for which semi-classical chronology protection theorems are of
dubious physical reliability. I consider a {\em ``Roman ring''} of
traversable wormholes, wherein a number of wormholes are arranged
in a ring in such a manner that no subset of wormholes is near to
chronology violation, though the combined system can be arbitrarily
close to chronology violation. I show that (with enough wormholes
in the ring) the gravitational vacuum polarization (the expectation
value of the quantum stress-energy tensor) can be made arbitrarily
small. In particular the back-reaction can be kept arbitrarily
small all the way to the ``reliability horizon''---so that
semi-classical quantum gravity becomes unreliable before the
gravitational back reaction becomes large.\\
\end{abstract}
\centerline{\em Accepted for publication in Physical Review D15.}
\twocolumn

\section{Introduction}

Working within the context of semi-classical quantum gravity,
Krasnikov~\cite{Krasnikov} and Sushkov~\cite{Sushkov} have recently
provided examples of two classes of spacetimes containing time
machines for which the gravitational vacuum polarization is
arbitrarily small all the way to the chronology horizon. In related
developments Kay, Radzikowski, and Wald~\cite{KRW}, and Cramer and
Kay~\cite{Kay-Cramer} have shown that these geometries suffer from
diseases on the chronology horizon itself. More recently, I have
argued~\cite{Reliability} that we should not physically trust
semi-classical quantum gravity once we reach the chronology horizon.

In this brief report I wish to present yet another class of spacetimes
for which the gravitational vacuum polarization can be made
arbitrarily small. Implications for chronology
protection~\cite{HawkingI,HawkingII} are briefly discussed.

\section{The Roman ring}

Given {\em one} wormhole, it appears (classically) to be absurdly
easy to turn it into a time
machine~\cite{Morris-Thorne,MTY,Kim-Thorne,Outrageous}, though
quantum effects vitiate this particular
approach~\cite{HawkingI,HawkingII,Kim-Thorne,Outrageous,Visser}.

Given {\em two} wormholes, it appears (even including quantum
effects) to be relatively easy to turn the compound system into a
time machine without each individual wormhole itself being a time
machine~\cite{Visser94,Lyutikov}.

Given {\em many} wormholes, I shall now show that it appears to be
even easier to turn the conglomeration into a time machine (with
all sub-collections of wormholes not themselves being time machines).

The key technical result is that, for any spacetime of non-trivial
topology, the gravitational vacuum polarization may be estimated
by adiabatic techniques to be

\begin{equation}
\langle T^{\mu\nu}(x) \rangle 
\approx \Delta_\gamma {\hbar\over s_\gamma(x,x)^4} t^{\mu\nu}.
\end{equation}

\noindent
Here $s_\gamma(x,x)$ is the length of the shortest spacelike geodesic
connecting the point $x$ to itself, while $t^{\mu\nu}$ is a
dimensionless tensor built out of the metric and tangent vectors
to this geodesic. $\Delta_\gamma$ is the van Vleck determinant
associated with this geodesic.

For the one-wormhole system this van Vleck determinant can be
estimated (insofar as the throat of the wormhole is reasonably
thin) to be close to $1$.  For the two-wormhole system the van
Vleck determinant is a complicated function of relative positions
and velocities. I will now provide a simple class of  multiple
wormhole configurations in which the van Vleck determinant is
calculable---and thereby show that there exists a class of geometries
for which the van Vleck determinant can be made arbitrarily small
all the way down to the chronology horizon.

\setlength{\unitlength}{1.0cm}
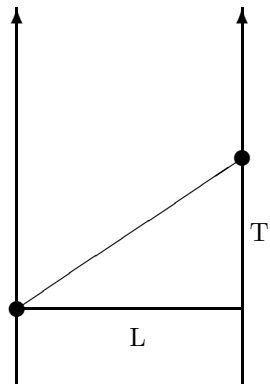
\begin{figure}[tb]
\begin{center}
\begin{picture}(5,5)(-1,0)
{
\thicklines
\put(0,0){\vector(0,1){5}}
\put(3,0){\vector(0,1){5}}
}
\put(0,1){\circle*{0.2}}
\put(3,3){\circle*{0.2}}
{
\thinlines
\put(0,1){\line(1,0){3}}
\put(0,1){\line(3,2){3}}
}
\put(1.5,0.50){L}
\put(3.1,1.9){T}
\end{picture}
\end{center}
\caption{Schematic representation of a single (chronology
respecting) wormhole. We identify two timelike lines that are
separated by a spatial jump $L$ and time shift $T$, with $T\ll L$.}
\end{figure}

Start by taking $N$ identical wormholes in otherwise flat Minkowski
spacetime. (These may be taken to be simple cut-and-paste wormholes
of the type discussed in~\cite{Visser,Visser89a,Visser89b}.) For
simplicity assume that all wormhole mouths are at rest with respect
to each other; so each wormhole is characterized by a spatial jump
$L$ and time-shift $T$ with $T \ll L$.

If we look at a geodesic that wraps once through a single wormhole,
the invariant interval is simply $s_1^2 = L^2 - T^2 \gg 0$.

Now arrange the $N$ wormholes in a big symmetric polygon, so that
the exit mouth of one wormhole is a normal-space distance of $\ell$
from the entrance mouth of the next wormhole.  (We will want $\ell
\ll L$, so that the normal space distance travelled to get from one
wormhole to the next is less than the distance then ``jumped'' by
going through the wormhole.)

\setlength{\unitlength}{1.0cm}
\begin{figure}[tb]
\begin{center}
\bigskip
\begin{picture}(5,5)
\put(0,1){\circle{0.5}}
\put(1,0){\circle{0.5}}
\put(0,4){\circle{0.5}}
\put(1,5){\circle{0.5}}
\put(4,0){\circle{0.5}}
\put(5,1){\circle{0.5}}
\put(4,5){\circle{0.5}}
\put(5,4){\circle{0.5}}
\put(0,1){\vector(+1,-1){1}}
\put(4,0){\vector(+1,+1){1}}
\put(5,4){\vector(-1,+1){1}}
\put(1,5){\vector(-1,-1){1}}
\put(0,0){$\ell$}
\put(0,5){$\ell$}
\put(5,0){$\ell$}
\put(5,5){$\ell$}
\put(+2.5,-0.0){L}
\put(+2.5,+5.0){L}
\put(-0.0,+2.5){L}
\put(+5.0,+2.5){L}
\end{picture}
\bigskip
\end{center}
\caption{Schematic representation of a Roman ring. This
example contains four wormholes. In each wormhole the two mouths
are separated by a spatial jump $L$. The normal space distance from
the exit mouth of one wormhole to the entrance mouth of the next
is $\ell$.}
\end{figure}
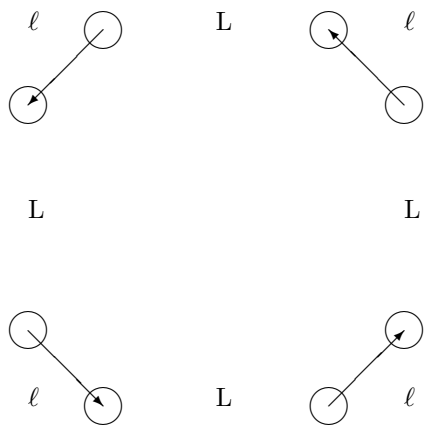

This ``Roman ring'' is a generalization of the two-wormhole ``Roman
configuration''~\cite{Morris-Thorne,Visser,Visser94,Lyutikov}.
(The realization that wormholes generically seem to imply time
travel can be traced back to an observation by Tom Roman~\cite{Outrageous}.
The ``Roman configuration'' was the first two-wormhole time machine
constructed by Morris and Thorne~\cite{Morris-Thorne}.) The
$N$-wormhole ``Roman ring'' can holistically be close to forming
a time machine even if its individual components are perfectly well
behaved.

Consider a closed geodesic that wraps once around the entire
compound system: this consists of $N$ segments of length $\ell$
in normal space, plus $N$ time-shifts of magnitude $T$ from going
through the $N$ wormholes. The invariant interval of this
once-through-the-system geodesic is

\begin{equation}
s_\gamma^2 = (N\ell)^2 - (N T)^2 = N^2 (\ell^2-T^2).
\end{equation}

\noindent
Because it is the normal-space distance between the wormhole mouths
that enters here ($\ell$; not the spatial jump $L$) we can easily
make this geodesic timelike.

For a conformally coupled massless field the symmetries enforce

\begin{equation}
t^{\mu\nu} = \eta^{\mu\nu} - 4 t^\mu t^\nu.
\end{equation}

\noindent
Here $\eta^{\mu\nu}$ is the spacetime metric, which is flat except
at the wormholes themselves; $t^\mu$ is the tangent vector to the
geodesic and is given (up to rotations) by

\begin{equation}
t^\mu 
= {(T,0,0,\ell)\over\sqrt{\ell^2-T^2}} 
= {N(T,0,0,\ell)\over s_\gamma}.
\end{equation}

\noindent
So anywhere along the geodesic $\gamma$ we can estimate

\begin{equation}
\langle T^{\mu\nu} \rangle 
\approx 
\Delta_\gamma {\hbar\over s_\gamma^4} 
\left[ 
\eta^{\mu\nu} - 4 {N^2 (T,0,0,\ell)^\mu (T,0,0,\ell)^\nu\over s_\gamma^2} 
\right].
\end{equation}

Of course, the only reason we are bothering with this Roman ring
geometry is because the extreme symmetry makes it relatively
easy to calculate the van Vleck determinant. Using the thin-throat
approximation, plus the tidal reformulation of the van Vleck
evolution equation presented in~\cite{Visser93}, we may calculate
the van Vleck determinant at the surface of any of the wormhole
throats to be~\cite{Visser94}

\begin{equation}
\Delta_\gamma = \left[ {N\over U_{N-1}(1+{\ell\over R})}\right]^2.
\end{equation}

\noindent
Here $U_N(x)$ is a Chebyshev polynomial of the second kind, and
$R$ is the radius of each wormhole mouth (assumed spherical). The
time shift $T$, and spatial jump $L$  quietly cancel out of the
calculation for the van Vleck determinant.

Proving the above result is a combinatoric agony that is presented
in excruciating detail in~\cite{Visser94}. That calculation was
carried out for a slightly different configuration: a geodesic that
wraps $N$ times through a single wormhole, but that calculation
can just as easily be adapted to the present case; a geodesic that
wraps once through $N$ wormholes---note that the high degree of
symmetry in the Roman ring configuration is essential for this
purpose.

If we are satisfied with the situation $\ell\gg R$ (a
perfectly sensible constraint---manipulating wormholes is likely to
be quite difficult enough without having them bump into each
other) then we can approximate

\begin{equation}
\Delta_\gamma \approx N^2 \left({R\over 2\ell}\right)^{2(N-1)}
\end{equation}

\noindent
As a consistency check, compare this with the results quoted by
Kim and Thorne~\cite{Kim-Thorne}, Lyutikov~\cite{Lyutikov}, and
Visser~\cite{Visser94}.

The physical interpretation for this result is simple: the van
Vleck determinant measures geometrically induced deviations from
the inverse-square law~\cite{Visser}. By assumption, we are sitting
right on top of one wormhole mouth, and the above result can be
thought of as due to $N-1$ defocussing events which occur as we
move through the $N-1$ other wormholes in the system to get back
to our starting point.

The van Vleck determinant may now be made as small as desired simply
by adding more wormholes to the system.

At the throat of any one of the wormholes, for any $\ell > T$, we
have

\begin{eqnarray}
\langle T^{\mu\nu} \rangle &\approx& 
\hbar\; {N^2\over s_\gamma^4} \; 
\left({R\over2\ell}\right)^{2(N-1)}
\nonumber\\
&&\times \left[ \eta^{\mu\nu} - 4 N^2 
{(T,0,0,\ell)^\mu \; (T,0,0,\ell)^\nu \over s_\gamma^2} \right].
\end{eqnarray}

\noindent
So at fixed $T$ and $\ell$ we have

\begin{eqnarray}
\langle T^{\mu\nu} \rangle &\approx& 
\hbar\; {1\over N^2(\ell^2-T^2)^2} 
\; \left({R\over2\ell}\right)^{2(N-1)} 
\nonumber\\
&&\times \left[ \eta^{\mu\nu} - 
4 {(T,0,0,\ell)^\mu \; (T,0,0,\ell)^\nu \over \ell^2 - T^2} \right].
\end{eqnarray}

\noindent
In particular $\langle T^{\mu\nu} \rangle \to 0$ as $N \to \infty$. 

Now suppose the whole system is adiabatically shrunk, keeping $T$
fixed but letting $\ell \to T^+$. The ``reliability
horizon''~\cite{Reliability}, the location at which we should
cease to believe the applicability of semi-classical quantum gravity,
will be located at $\sqrt{\ell^2 - T^2} = \ell_{Planck}$. 

(General arguments supporting this designation are provided
in~\cite{Reliability}. In the present more specific context it
suffices to realise that once $\sqrt{\ell^2 - T^2} < \ell_{Planck}$,
any quantum field [including gravitons] propagating on this background
will be subject to Planck scale physics.)

At the throat of any one wormhole, when the system is at the
reliability horizon, we have

\begin{eqnarray}
\langle T^{\mu\nu} \rangle &\approx& 
\hbar\; {1\over N^2 \ell_{Planck}^4} 
\; \left({R\over2\ell}\right)^{2(N-1)} 
\nonumber\\
&&\times \left[ \eta^{\mu\nu} - 
4 {(\ell,0,0,\ell)^\mu \; (\ell,0,0,\ell)^\nu \over \ell_{Planck}^2} \right].
\end{eqnarray}

\noindent
Again $\langle T^{\mu\nu} \rangle \to 0$ as $N \to \infty$. 

Thus with enough wormholes, we can arrange the gravitational vacuum
polarization,  and therefore the back-reaction, to be arbitrarily
small all the way down to the reliability horizon.

\section{Implications}

This counter-example is enough to show that it is impossible to
come up with a chronology protection theorem that makes reference
only to the ``reliable region''---and so it is impossible to come
up with chronology protection theorem that is {\em physically
reliable} within the context of semi-classical quantum gravity.

In this regard I am completely in agreement with Krasnikov~\cite{Krasnikov}
and Sushkov~\cite{Sushkov}, though the current class of models is
obtained in a radically different (and perhaps more physically
transparent) manner.

My interpretation is perhaps a little different: I view this not
as a vindication for time travel enthusiasts but rather as an
indication that resolving issues of chronology protection requires
a fully developed theory of quantum gravity~\cite{Reliability}.

\acknowledgements

This research was supported by the U.S. Department of Energy.  I
wish to thank Tom Roman for reading the manuscript and providing
helpful comments.


\end{document}